\documentclass[12pt]{article}
\usepackage{geometry}
\usepackage{amsmath}
\usepackage{amssymb}
\usepackage{amsmath}
\usepackage{graphicx}
\numberwithin{equation}{section}

\def\appendix#1{
  \addtocounter{section}{1}
  \setcounter{equation}{0}
  \renewcommand{\thesection}{\Alph{section}}
 \section*{Appendix \thesection\protect\indent \parbox[t]{11.715cm} {#1}}
  \addcontentsline{toc}{section}{Appendix \thesection\ \ \ #1}
  }

\renewcommand{\thefootnote}{\fnsymbol{footnote}}

\newcommand{\ff}{h(\lambda)}
\newcommand{\comment}[1]{}
\newcommand{\be}{\begin{equation}}
\newcommand{\ee}{\end{equation}}
\newcommand{\ba}{\begin{aligned}}
\newcommand{\ea}{\end{aligned}}

\def\m1{\left(-1\right)^{F_i}}

\makeatletter
\def\sla@#1#2#3#4#5{{%
  \setbox\z@\hbox{$\m@th#4#5$}%
  \setbox\tw@\hbox{$\m@th#4#1$}%
  \dimen4\wd\ifdim\wd\z@<\wd\tw@\tw@\else\z@\fi
  \dimen@\ht\tw@
  \advance\dimen@-\dp\tw@
  \advance\dimen@-\ht\z@
  \advance\dimen@\dp\z@
  \divide\dimen@\tw@
  \advance\dimen@-#3\ht\tw@
  \advance\dimen@-#3\dp\tw@
  \dimen@ii#2\wd\z@  \raise-\dimen@\hbox to\dimen4{%
    \hss\kern\dimen@ii\box\tw@\kern-\dimen@ii\hss}%
  \llap{\hbox to\dimen4{\hss\box\z@\hss}}}}
\def\slashed#1{%
  \expandafter\ifx\csname sla@\string#1\endcsname\relax
    {\mathpalette{\sla@/00}{#1}}%
  \else
    \csname sla@\string#1\endcsname
  \fi}
\makeatother

\newcommand{\beq}{\begin{equation}}
\newcommand{\eeq}{\end{equation}}
\newcommand\beqa{\begin{eqnarray}}
\newcommand\eeqa{\end{eqnarray}}
\newcommand\bea{\begin{array}}
\newcommand\eea{\end{array}}

\newcommand{\nn}{\nonumber}
\newcommand{\neqa}{\nonumber\end{eqnarray}}
\newcommand{\la}{\label}

\newcommand{\OO}{{\cal O}}
\newcommand{\color}[1]{}

\newcommand{\eq}[1]{(\ref{#1})}

\renewcommand{\t}{\tilde}

\def\({\left(}
\def\){\right)}
\def\[{\left[}
\def\]{\right]}

\def\<{\langle}
\def\>{\rangle}

\def\d{\partial}

\begin{document}

%%%%%%%%%%%%%%%%%%%%%%%%%%%%%%%%%%%%%
%%%%%%%%%%%%%%%%%%%%%%%%%%%%%%%%%%%%%
%%%%%%%%%%%%%%%%%%%%%%%%%%%%%%%%%%%%%

\thispagestyle{empty}
\begin{flushright}\footnotesize
%\texttt{arxiv:yymm.nnnn}\\
%\texttt{CALT-68-NNNN}\\
\texttt{LPTENS 08/39}\\
\texttt{SPhT-t08/NNN}\\
\vspace{2.1cm}
\end{flushright}

\renewcommand{\thefootnote}{\fnsymbol{footnote}}
\setcounter{footnote}{0}
\setcounter{figure}{0}
\begin{center}
{\Large\textbf{\mathversion{bold} The all loop $AdS_4/CFT_3$ Bethe ansatz}\par}

\vspace{2.1cm}

\textrm{Nikolay Gromov$^{\alpha}$,  Pedro Vieira$^{\gamma}$}
\vspace{1.2cm}

\textit{$^{\alpha}$ Service de Physique Th\'eorique,
CNRS-URA 2306 C.E.A.-Saclay, F-91191 Gif-sur-Yvette, France;
Laboratoire de Physique Th\'eorique de
l'Ecole Normale Sup\'erieure et l'Universit\'e Paris-VI,
Paris, 75231, France;
St.Petersburg INP, Gatchina, 188 300, St.Petersburg, Russia } \\
\texttt{nikgromov@gmail.com}
\vspace{3mm}

\textit{$^{\gamma}$ Laboratoire de Physique Th\'eorique
de l'Ecole Normale Sup\'erieure et l'Universit\'e Paris-VI, Paris,
75231, France;  Departamento de F\'\i sica e Centro de F\'\i sica do
Porto Faculdade de Ci\^encias da Universidade do Porto Rua do Campo
Alegre, 687, \,4169-007 Porto, Portugal} \\
\texttt{pedrogvieira@gmail.com}
\vspace{3mm}

%%%%%%%%

\par\vspace{1cm}

\textbf{Abstract}\vspace{5mm}
\end{center}

\noindent
We propose a set of Bethe equations yielding the full asymptotic spectrum of the $AdS_4/CFT_3$ duality proposed in arXiv:0806.1218 to all orders in the t'Hooft coupling. These equations interpolate between the 2-loop Bethe ansatz of Minahan and Zarembo arXiv:0806.3951 and the string algebraic curve of arXiv:0807.0437. The several $SU(2|2)$ symmetries of the theory seem to highly constrain the form of the Bethe equations up to a dressing factor whose form we also conjecture.
\vspace*{\fill}

\setcounter{page}{1}
\renewcommand{\thefootnote}{\arabic{footnote}}
\setcounter{footnote}{0}

\newpage

%%%%%%%%%%%%%%%%%%%%%%%%%%%%%%%%%%%%%
%%%%%%%%%%%%%%%%%%%%%%%%%%%%%%%%%%%%%
%%%%%%%%%%%%%%%%%%%%%%%%%%%%%%%%%%%%%

%\tableofcontents
%\newpage
%%%%%%%%%%%%%%%%%%%%%%%%%%%%%%%%%%%%%
%%%%%%%%%%%%%%%%%%%%%%%%%%%%%%%%%%%%%
%%%%%%%%%%%%%%%%%%%%%%%%%%%%%%%%%%%%%

\section{Conjecture and discussion}

A fascinating $AdS_4/CFT_3$ duality was recently proposed by Aharony, Bergman, Jafferis and Maldacena in \cite{Aharony:2008ug} following previous interesting works \cite{Schwarz:2004yj} (for subsequent developments see \cite{pos}). According to this duality the large $N$ limit of a particular  three dimensional superconformal $SU(N)\times SU(N)$ Chern-Simons theory with level $k$ is dual to  M-theory on $AdS_4\times S^7/Z_k$. Furthermore, when we take the limit $k,N\to \infty$, with the t'Hooft coupling
\beq
\lambda=N/k \equiv 8 g^2
\eeq
held fixed, we obtain a remarkable correspondence between a planar gauge theory and free type IIA superstring theory in $AdS_4\times CP^3$. We will always work in this limit.

In the beautiful paper by Minahan and Zarembo \cite{MZ} the $SU(4)$ sector of the Chern-Simons theory was shown to be integrable to 2 loops in perturbation theory, the leading order for this model. The $OSp(2,2|6)$ nested Bethe equations yielding the complete spectrum of all single trace operators to $2$-loops were also proposed in this paper. At strong coupling, when the theory can be described by a supercoset sigma model, integrability was shown in \cite{today1,today2} and the finite gap construction \cite{C1,C2,C3,C4,C5,C6} of the algebraic curve was carried in  \cite{curve}. Here we propose a set of five Bethe equations yielding the spectrum of the theory for any value of the t'Hooft coupling.

In the context of the $AdS_5/CFT_4$ such equations were proposed by Beisert and Staudacher \cite{BS}. In their proposal solely a scalar factor was unfixed. This factor was latter conjectured by Beisert, Eden and Staudacher in \cite{BES}. For an incomplete list of references on the topic of integrability in $AdS/CFT$ see \cite{MZ2,huge} and the several other references in this manuscript. The Beisert-Staudacher equations are asymptotic and valid only for large operators \cite{Mathias}.  Our equations are also asymptotic and do not take into account wrapping interactions. It would be extremely interesting to investigate these type of corrections here  (see \cite{wrap1,wrap2} for related papers on such effects in the $AdS_5\times CFT_4$ duality).

We will now present our conjecture. In the end of this section we shall mention the three main arguments
supporting it and in sections \ref{sec1}, \ref{sec2} and \ref{sec3} we expand on each of these points.

We start by defining the usual Zhukowsky variables
$$
x+\frac{1}{x}=
\frac{u}{\ff}\;\;,\;\;x^\pm+\frac{1}{x^\pm}=\frac{1}{\ff}\(u\pm\frac
i2\) \,.
$$
where $\ff$ is a yet to be fixed function introduced in \cite{alsof1,BMN2,BMN3}\footnote{We use $\ff$ instead of $f(\lambda)$ because we will save the latter for the scaling function discussed below.}. It interpolates between \cite{alsof1,BMN2,BMN3}
%\footnote{The simplest function with these asymptotics is $\ff=\sqrt{\frac{\lambda^2}{2\lambda+1}}$. It has branch cut on the complex plane of %$\lambda$. It would be interesting to investigate the meaning of  singularities of $\ff$.}
\beqa
\ff\simeq \lambda\,\,\,\,\, \text{, at weak coupling}\la{f1}
\eeqa
and
\beqa
\ff\simeq \sqrt{\lambda/2}\,\,\,\,\, \text{, for large values of the t'Hooft coupling.} \la{f2}
\eeqa
In the end of this section we shall provide some speculative comments about this function.

\begin{figure}[t]
\centering \resizebox{90mm}{!}{\includegraphics{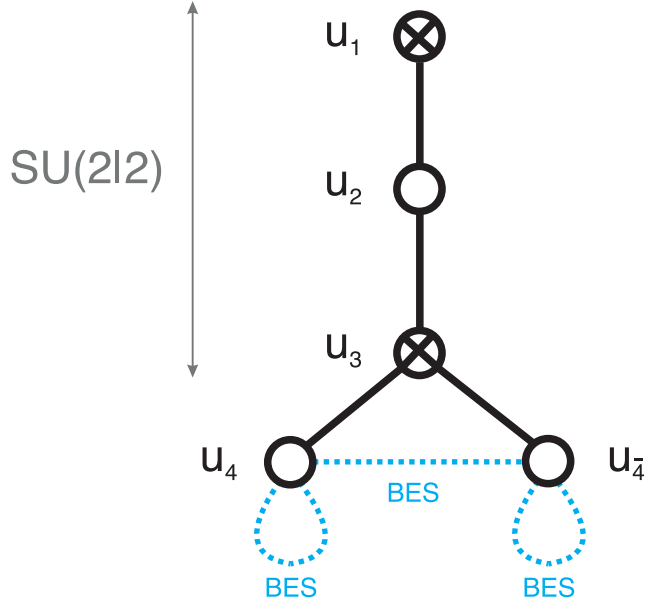}}
\caption{\small  Structure of the $AdS_4/CFT_3$ Bethe equations. The several Bethe equations are organized according to the $OSp(2,2|6)$ symmetry of the problem. The Dynkin diagram associated with this symmetry group is depicted in the figure. Particularly important subsectors are two $SU(2|2)$ obtained by exciting solely momentum carrying roots of one of the wings ($u_4$ or $u_{\bar 4}$) plus an arbitrary amount of auxiliary roots in the $SU(2|2)$ tail ($u_1$, $u_2$ and $u_3$). Equally important is the $SU(2)\times SU(2)$ subsector obtained by only exciting the momentum carrying roots ($u_4$ and $u_{\bar 4}$). When we consider higher orders in perturbation theories a dressing kernel appears introducing extra self-interactions for the momentum carrying roots and also a new interaction between the roots $u_4$ and $u_{\bar 4}$. Perturbatively, this couples the two $SU(2)$'s in the $SU(2)\times SU(2)$ sector starting at eight loops.} \label{BAEpic}
\end{figure}

Next we introduce five types of Bethe roots $ u_1,u_2,u_3,u_4$ and $ u_{\bar{4}} $. The spectrum of all conserved charges is then given by the momentum carrying roots $u_4$ and $u_{\bar 4}$ alone from
\beq
{\cal Q}_n= \sum_{j=1}^{K_4} \textbf{q}_n(u_{4,j})+ \sum_{j=1}^{K_4} \textbf{q}_n(u_{\bar 4,j}) \,\,\, , \,\,\, \textbf{q}_n=\frac{i}{n-1} \(\frac{1}{(x^+)^{n-1}}-\frac{1}{(x^-)^{n-1}}\) \la{charges}
\eeq
%\beqa
%E=\frac{\ff}{4i}  \sum_{j=1}^{K_4} \(x_{4,j}^+-\frac{1}{x_{4,j}^+}-x^-_{4,j}+\frac{1}{x^-_{4,j}}\) +\frac{\ff}{4i}  \sum_{j=1}^{K_{\bar 4}} \(x^+_{\bar 4,j}-\frac{1}{x^+_{\bar 4,j}}-x^-_{\bar 4,j}+\frac{1}{x^-_{\bar 4,j}}\)
%\eeqa
and the spectrum of anomalous dimensions (or string states energies) follows from
\beqa
E =\ff{\cal Q}_2 \,.\la{eq:E}
\eeqa
In terms of  $p_j=\frac{1}{i}\log\frac{x^+_{4,j}}{x^-_{4,j}}$ and $\bar p_j=\frac{1}{i}\log\frac{x^+_{\bar 4,j}}{x^-_{\bar 4,j}}$, we have
\beqa
E =\sum_{j=1}^{K_4} \frac{1}{2}\(\sqrt{1+16 \ff^2 \sin^2 \frac{p_j}{2}} -1\)+\sum_{j=1}^{K_{\bar 4}} \frac{1}{2}\(\sqrt{1+16 \ff^2 \sin^2 \frac{\bar p_j}{2}} -1\) \,. \la{Ep}
\eeqa
Finally, and most importantly, the Bethe roots are quantized through the Bethe equations
\begin{eqnarray}
1&=&
\prod_{j=1}^{K_2}
\frac{u_{1,k}-u_{2,j}+\frac{i}{2} }{u_{1,k}-u_{2,j}-\frac{i}{2} }
\prod_{j=1}^{K_{4}}
\frac{1-1/x_{1,k} x^+_{4,j}}{1-1/x_{1,k}x_{4,j}^-}
\prod_{j=1}^{K_{\bar 4}}
\frac{1-1/x_{1,k} x^+_{\bar 4,j}}{1-1/x_{1,k}x_{\bar 4,j}^-} \,,
 \nn \\
\nn 1&=&
\prod_{j\neq k}^{K_2}
\frac{u_{2,k}-u_{2,j}-i }{u_{2,k}-u_{2,j}+i }
\prod_{j=1}^{K_1}
\frac{u_{2,k}-u_{1,j}+\frac{i}{2} }{u_{2,k}-u_{1,j}-\frac{i}{2} }
\prod_{j=1}^{K_3}
\frac{u_{1,k}-u_{3,j}+\frac{i}{2} }{u_{1,k}-u_{3,j}-\frac{i}{2} }\,,
\\
\nn 1&=&
\prod_{j=1}^{K_2}
\frac{u_{3,k}-u_{2,j}+\frac{i}{2} }{u_{3,k}-u_{2,j}-\frac{i}{2} }
\prod_{j=1}^{K_4}
\frac{x_{3,k} -x^+_{4,j}}{x_{3,k} -x^-_{4,j}}
\prod_{j=1}^{K_{\bar 4}}
\frac{x_{3,k} -x^+_{\bar 4,j}}{x_{3,k} -x^-_{\bar 4,j}} \,
\\
\(\frac{x^+_{4,k}}{x^-_{4,k}}\)^{
L} \la{BAE}&=&
\prod_{j\neq k}^{K_4}
\frac{u_{4,k}-u_{4,j}+i}{u_{4,k}-u_{4,j}-i}  \,
\prod_{j=1}^{K_1}
\frac{1-1/x^-_{4,k} x_{1,j}}{1-1/x^+_{4,k} x_{1,j}}
\prod_{j=1}^{K_3}
\frac{x^-_{4,k}-x_{3,j} }{x^+_{4,k}-x_{3,j}}   \times
 \\
\nn &\times &\prod_{j=1}^{K_4}
\sigma_{\rm BES}(u_{ 4,k},u_{ 4,j}) \prod_{j=1}^{K_{\bar 4}} \sigma_{\rm BES}(u_{ 4,k},u_{ \bar 4,j})  \,, \\
\(\frac{x^+_{\bar 4,k}}{x^-_{\bar  4,k}}\)^{
L} &=&
\prod_{j=1}^{K_{\bar 4}}
\frac{u_{\bar  4,k}-u_{\bar 4,j}+i}{u_{\bar 4,k}-u_{\bar 4,j}-i}  \,
\prod_{j=1}^{K_1}
\frac{1-1/x^-_{\bar 4,k} x_{1,j}}{1-1/x^+_{\bar 4,k} x_{1,j}}
\prod_{j=1}^{K_3}
\frac{x^-_{\bar 4,k}-x_{3,j} }{x^+_{\bar 4,k}-x_{3,j}}   \times
\nn \\
&\times &\prod_{j\neq k}^{K_{\bar 4}}
\sigma_{\rm BES}(u_{ \bar 4,k},u_{ \bar 4,j}) \prod_{j=1}^{K_{  4}} \sigma_{\rm BES}(u_{ \bar 4,k},u_{ 4,j}) \nn \,.
\end{eqnarray}
The number of roots is related to the Dynkin labels of the state as in \cite{MZ}.
Furthermore we must consider only solutions subject to the zero momentum condition \cite{MZ}
\beqa
1= \prod_{j=1}^{K_4} \frac{x^+_{4,j}}{x^-_{4,j}} \prod_{j=1}^{K_{\bar 4}} \frac{x^+_{\bar 4,j}}{x^-_{\bar 4,j}} \;\;\Leftrightarrow\;\;{\cal Q}_1=2\pi m\,.\la{mom}
\eeqa
The structure of the nested Bethe equations can be represented in figure \ref{BAEpic}.
The function $\sigma$ is the so called dressing factor. It is given precisely by the same form as in the work of Beisert, Eden and Staudacher except that there one had a dressing kernel $\sigma$ given by
\beq
\sigma=\sigma_{\rm BES}^2
\eeq
manifesting the $SU(2|2) \otimes SU(2|2)$ symmetry of the problem whereas here we have
\beq
\sigma=\sigma_{\rm BES}
\eeq
appearing in two equations as a consequence of the $SU(2|2)_A\oplus SU(2|2)_B$ symmetry. Furthermore the coupling $g_{{\rm AdS_5}}=\frac{\sqrt{\lambda_{\rm AdS_5}}}{4\pi}$ in the BES kernel should be now replaced by $\ff$.
The dressing kernel can be written in a simple integral form as \cite{DHM}
\beq
\sigma_{\rm BES}(u_j,u_k)=e^{i\theta_{jk}}\;\;,\;\;
\theta_{jk}={\chi(x_j^+,x_k^+)+ \chi(x_j^-,x_k^-)- \chi(x_j^+,x_k^-)- \chi(x_j^-,x_k^+) -(k\leftrightarrow j)}
\eeq
with
\beq
\chi(x,y)=-i\oint \frac{dz_1}{2\pi}\oint \frac{dz_2}{2\pi} \frac{1}{(x-z_1)(y-z_2)} \log\Gamma(1+i \ff (z_1+1/z_1-z_2-1/z_2))
\eeq
integrated over the contours $|z_1|=|z_2|=1$. This kernel interpolates between
\beq
\displaystyle\sigma_{\rm BES}(u,v) \stackrel{\lambda\to 0}{\to} 1\,,
\eeq
at weak coupling and
\beq
\sigma_{\rm BES}(u_j,u_k) \stackrel{\lambda\to \infty}\to
 \frac{1-1/x_k^+ x_j^-}{1-1/x_k^-x_j^+}  \(\frac{x_k^-
x_j^--1}{x_k^-x_j^+-1} \frac{x_k^+ x_j^+-1}{x_k^+ x_j^--1}\)^{ i(u_k-u_j)} \equiv \sigma_{{\rm AFS}}(u_j,u_k)\,, \la{AFSl}
\eeq
for large values of the t'Hooft coupling.  $\sigma_{\rm AFS}$ is the AFS dressing kernel proposed in \cite{AFS} in the study of the quantum string Bethe equations for the $AdS_5\times S^5$ string.

There are three main reasons to believe the correctness of our proposal:
\begin{enumerate}
\item{It exhibits an $OSp(2,2|6)$ global symmetry, matches the 2-loop equations of Minahan and Zarembo \cite{MZ}, and possesses a nontrivial set of Weyl dualities which probably ought to be present for the correct set of Bethe equations. This will be further explained in section \ref{sec1}. In appendix A we present (\ref{BAE}) in the two possible gradings obtained by applying the fermionic dualities to our equations.}
\item{It yields the algebraic curve of \cite{curve} in the continuum limit at strong t'Hooft coupling. Therefore it encapsulates all superstring classical dynamics. Moreover the scalar factor plus the finite size corrections are constrained by the string  semi-classical quantization in a very natural way. This point will be discussed in section \ref{sec2}. }
\item{Finally, it seems that the form of these equations is highly constrained, if not fixed, by the $SU(2|2)$ symmetry of the problem as discussed in section \ref{sec3}. This again mimics the past developments in $\mathcal{N}=4$ SYM where it turned out to be case \cite{B22} (see also \cite{Arutyunov:2006yd}).
}
\end{enumerate}
In the next three sections we shall develop on each of these points. In the remain of this section let us comment on some curious features of our proposal and mention some future work proposals.

The BES kernel can be written in several ways. Above we used the integral representation of Dorey, Hofman and Maldacena \cite{DHM}. Another useful writing of the BES kernel in terms of the charges introduced above (\ref{charges}) is
\beq
\sigma(u_j,u_k)=e^{i\theta_{jk}}\;\;,\;\;\theta_{jk}=\sum_{r=2,s=r+1}c_{r,s}\[\textbf{q}_r(x_j)\textbf{q}_s(x_k)-\textbf{q}_r(x_k)\textbf{q}_s(x_j)\]
\eeq
where the coefficients $c_{r,s}$ are given in \cite{BES}
\beq
c_{r,s}=\ff\delta_{r+1,s}+\frac{1+(-1)^{r+s}}{\pi}\frac{(r-1)(s-1)}{(r+s-2)(s-r)}+\OO\(1/\sqrt\lambda\)\;. \la{crs}
\eeq
The leading order yields the AFS phase \cite{AFS} and the next to leading order produces the HL factor \cite{HL}.
Notice furthermore that the products of the BES kernels in (\ref{BAE}) can be written as \cite{AFS}
\beq
\prod_{j=1}^{K_4} \sigma_{\rm BES}(u_{4,k},u_{4,j}) \prod_{j=1}^{K_{\bar 4} }\sigma_{\rm BES}(u_{4,k},u_{\bar 4,j})=  \exp \(\, \sum_{r=2,s=r+1} i\, c_{r,s}\,\(\textbf{q}_r(x_{4,k}) \mathcal{Q}_s-\textbf{q}_s(x_{4,k})\mathcal{Q}_r\) \) \,, \la{HLQ}
\eeq
with a similar expression when $u_{4,k}$ is replaced by $u_{\bar 4,k}$. Written in this form, the dressing kernel looks precisely like the one appearing in the $AdS_5/CFT_4$ duality \cite{BES} apart from a factor of $2$ and the replacement of $g_{\rm AdS_5}=\frac{\sqrt{\lambda_{\rm AdS_5}}}{4\pi}$ by $\ff$. The interaction between the $u_{4}$ and the $u_{\bar 4}$ appears because the charges (\ref{charges}) are the sum of the charges of the $u_{4}$ roots with the charges of the $u_{\bar 4}$ roots. There is strong evidence that this writing of dressing kernels is fairly generic \cite{Beisert:2005wv}.

So, assuming our proposal to be correct there are many interesting projects to be addressed. First of all, the amount  of intermediate steps leading to the Beisert-Staudacher equations \cite{BS} since the seminal paper \cite{MZ2} was quite significative \cite{huge,C1,C2,C3,C4,C5,C6,Mathias}. They should be taken again for this new theory. For example,  it would be interesting to understand crossing symmetry for this model as done in \cite{Janik} for the $AdS_5/CFT_4$ duality.

One should also understand the spectrum of bound states and relate it with the several singularities of the $S$-matrix as done in \cite{nick,DHM}\footnote{We thank N.Dorey for discussions and explanations on this topic.}. We notice that when we consider each $SU(2|2)$ sector separately the double poles coming from the BES kernel found in \cite{DHM} become single poles. This could be related to the crossing transformation mention above. Indeed, it was seen in \cite{DHM} that to assemble the several three point vertices into box diagrams, crossed vertices were often important and those seem to relate the two $SU(2|2)$ present in the $AdS_5/CFT_4$ chain. On the other hand there are eight fluctuations which from the Bethe ansatz point of view require us to consider stacks \cite{C6} of bound states with one $u_4$ and one $u_{\bar 4}$ root. For those, when fusing the Bethe equations in the usual way, by multiplying the equations for roots $u_4$ and $u_{\bar 4}$, we will obtain again $\sigma_{BES}^2$ and thus the double poles reappear consistently with the remark just done. It would be very important to understand these points in greater detail.

We are aware that our conjecture is based on a very limited amount of data and of course we are relying a lot on the experience acquired with the remarkable developments in the $AdS_5/CFT_4$ duality.

Another major investigation subject that must be tackled seriously is the study of the wrapping interactions mentioned above. After all, at the end of the day we want to compute the spectrum of simple and small operators for any values of the t'Hooft coupling.

Finally, a more immediate and permeant problem would be to compute the interpolating function $\ff$ which appears in the magnon dispersion relation,
\beq
\epsilon(p) \sim \sqrt{1+16 \ff^2\sin^2\frac{p}{2} }\,. \la{sqrt}
\eeq
In $\mathcal{N}=4$ SYM this function is believed to be simply $16 \ff^2=\lambda_{\rm AdS_5}/\pi$ for all values of the t'Hooft coupling. In our case we know the function is more complicated but still, given the simplicity we observe in the $AdS_5/CFT_4$ duality, it is reasonable to expect $\ff$ to be given by some simple expression. We know that the small $\lambda$ expansion of $\ff^2$ contains only even powers because the perturbative Chern-Simons theory is organized in this way. On the other hand we know that $\ff^2 \sim \lambda$ for large values of the t'Hooft coupling (\ref{f2}). This means the function should have a square root cut in the complex plane. An example of such function compatible with the asymptotics (\ref{f1}) and (\ref{f2}) is\footnote{Among some other proposals, this $h(\lambda)$ was also written in \cite{alsof1}. We thank Tadashi Takayanagi for pointing this out to us.
}
\beq
\ff^2={ \lambda^2 \over \sqrt{1+4\lambda^2}} \,.
\eeq
This function behaves as
\beq
\ff=\sqrt{{\lambda \over 2}} +\mathcal{O}(1/\sqrt{\lambda})
\eeq
so that the $\mathcal{O}(1)$ term is not present. This is actually a property we must require for the interpolating function because the leading term in (\ref{crs}) -- giving the AFS phase -- should not mix with the subleading contribution -- yielding the HL factor --  otherwise semi-classics will not work as we explain in section \ref{HL}. To check this necessary behavior we can compute the one-loop shift around the giant magnon solution \cite{HM,BMN2,BMN3} and verify that it vanishes \cite{igor}.
We also point out the curious property of the simple function we wrote down, relating its weak and strong coupling expansions:
\beqa
\ff=\sum_{n=1}^{\infty} c_n \(2 \lambda\)^{2n} =\sum_{n=1}^{\infty} \t c_n \(2\lambda\)^{3-2n}
\eeqa
where
\beqa
c_n=\t c_n \,.
\eeqa
This kind of analytical behavior was seen to be present in the dressing kernel in \cite{BES}. Here however, it is so far a simple curiosity.

Another important hint would be to analyze the radius of convergence of the planar Chern-Simons theory. For example, in the $\mathcal{N}=4$ SYM theory we find that the values of $\lambda$ such that the argument of the square root (\ref{sqrt}) first vanishes are at $|\lambda|=\pi^2$ (for $p=\pi$) which corresponds to the radius of convergence of the gauge theory.

It would be extremely interesting to find the precise value of $\ff$.

To check our conjecture it is useful to find good quantities with a smooth interpolation from the weak to strong coupling limit. One such quantity is the (generalized) scaling function\footnote{The results concerning the scaling function presented in this paper benefited largely from discussion with D.Serban and D.Volin whom we thank.} $f(\lambda)$ much studied in the $AdS_5/CFT_4$ duality (see for example \cite{flambda,ES,BES,FRS} and references therein). In the recent paper \cite{FRS} a review and definition of this quantity is provided. It is quite an interesting quantity because it can be accessed from field theoretical computations to very high loop order \cite{weakf,strongf}. It follows from our conjectured Bethe equations that this function can be trivially obtained from the analogous quantity in $\mathcal{N}=4$ SYM from\footnote{The generalized scaling function can also be easily obtained from the $\mathcal{N}=4$ SYM result as $$f_{CS}\(\lambda,\frac{J}{\log(S)}\)=\frac{1}{2}\,f_{N=4}\(\lambda,\frac{2J}{\log(S)}\)_{\frac{\sqrt{\lambda}}{4\pi}\to h(\lambda) }\,.$$.}
\beq
f_{CS}(\lambda)=\frac{1}{2} f_{\mathcal{N}=4}(\lambda) \Big|_{\frac{\sqrt{\lambda}}{4\pi}\to \ff} \la{hCS} \,.
\eeq
This map  follows from the observation, explained in appendix A.1, that the BES equations \cite{BES} (and even the FRS equations \cite{FRS})  are exactly the same for the Chern-Simons theory provided we replace the coupling constant as in (\ref{hCS})! Therefore, in particular,
\beq
f_{CS}(\lambda) = 4 h^2(\lambda)-\frac{4}{3}\pi^2 h^4(\lambda)+\frac{44}{45}\pi^4 h^6(\lambda)+\dots = 4\lambda^2+\mathcal{O}(\lambda^4) \,\,\, ,\,\,\, \lambda\ll 1
\eeq
at weak coupling and
\beq
f_{CS}(\lambda) = 2 \ff-\frac{3\log2}{2\pi}-\frac{K}{8\pi^2}\frac{1}{\ff}+ \dots=\sqrt{2 \lambda} -\frac{3\log2}{2\pi} +\mathcal{O}(1/\sqrt{\lambda}) \,\,\, ,\,\,\, \lambda\gg 1
\eeq
at strong coupling.

The leading weak coupling value can be computed relying solely on the Minahan-Zarembo Bethe equations, see appendix A.1, without ever using our conjectured equations. This term was also identified in \cite{Aharony:2008ug} (formula (4.17)). There is a factor of $4$ which does not seem to match and which ought to be understood. We comment further on this point in appendix A.1.

Since the strong coupling asymptotics can be computed from the two leading coefficients of the dressing factor alone -- which are derived in section \ref{sec2} --  they should certainly be right.  The leading strong coupling coefficient was also  identified in \cite{Aharony:2008ug} (formula (4.16)). We find a precise agreement with their prediction.

In the next three sections we develop each of the three topics providing evidence for (\ref{BAE}) which we mentioned above.

\section{Weak coupling limit and dualities in Bethe equations} \la{sec1}
In the weak coupling limit we have
\beq
x^\pm \rightarrow \frac{u\pm i/2}{ \ff}  \,\, , \,\, x\to   \frac{u}{ \ff}
\eeq
with $\ff\to 0$.  Therefore the proposed Bethe equations (\ref{BAE}) simplify dramatically to
\beqa\label{betheeqscfalt}
1&=&\prod_{k=1}^{{K_{2}}}\frac{u_{1,j}-u_{2,k}+i/2}{u_{1,j}-u_{2,k}-i/2}\,, \\
1&=& \prod_{k\neq j}^{{K_{2}}}\frac{u_{2,j}-u_{2,k}-i}{u_{2,j}-u_{2,k}+i}\prod_{k=1}^{{K_{3}}}\frac{u_{2,j}-u_{3,k}+i/2}{u_{2,j}-u_{3,k}-i/2}\prod_{k=1}^{{K_{1}}}\frac{u_{2,j}-u_{1,k}+i/2}{u_{2,j}-u_{1,k}-i/2}\nn\\
1&=&\prod_{k=1}^{{K_{4}}}\frac{u_{3,j}-u_{4,k}-i/2}{u_{3,j}-u_{4,k}+i/2}\prod_{k=1}^{{K_{\bar 4}}}\frac{u_{3,j}-u_{\bar 4,k}-i/2}{u_{3,j}-u_{\bar 4,k}+i/2}\prod_{k=1}^{{K_{2}}}\frac{u_{3,j}-u_{2,k}+i/2}{u_{3,j}-u_{2,k}-i/2}\nn\\
\left(\frac{u_{4,j}+i/2}{u_{4,j}-i/2}\right)^L&=&\prod_{k\neq j}^{{K_{4}}}\frac{u_{4,j}-u_{4,k}+i}{u_{4,j}-u_{4,k}-i} \prod_{k=1}^{{K_{3}}}\frac{u_{4,j}-u_{3,k}-i/2}{u_{4,j}-u_{3,k}+i/2}\nn\\
\left(\frac{u_{\bar 4,j}+i/2}{u_{\bar 4,j}-i/2}\right)^L&=&\prod_{k\neq
j}^{{K_{\bar 4}}}\frac{u_{\bar 4,j}-u_{\bar 4,k}+i}{u_{\bar 4,j}-u_{\bar 4,k}-i}
\prod_{k=1}^{{K_{3}}}\frac{u_{\bar 4,j}-u_{3,k}-i/2}{u_{\bar 4,j}-u_{3,k}+i/2}\nn
\eeqa
which are precisely the $1$-loop Bethe equations of Minahan and Zarembo \cite{MZ}.
The energy becomes
\beqa
E =\ff{\cal Q}_2\simeq \ff^2\(\sum_{i}^{K_u}\frac{1}{u_{4,j}^2+1/4}+\sum_{i}^{K_v}\frac{1}{u_{\bar 4,j}^2+1/4}\) \,,\la{eq:E0}
\eeqa
which perfectly matches with \cite{MZ} with $\ff\simeq \lambda$ at weak coupling.

This structure of the Bethe equations was guessed \cite{MZ} based on the $OSp(2,2|6)$ symmetry of the problem. Bethe roots associated with bosonic nodes in the Dynkin diagram self-interact and interact with the neighbors in the Dynkin diagram. The same holds for fermionic roots except for the self interaction which is absent. The relative factors of $\pm 2$ or $\pm 1$ before the several $i/2$ in these equations are precisely predicted by the $OSp(2,2|6)$ Dynkin diagram.

When generalizing to the all loop case we should keep the $OSp(2,2|6)$ symmetry. In particular there exists a plethora of dualities in Bethe ansatz which transform various configurations of Bethe roots into some other configurations \cite{Tsuboi,C6,GV3}. These transformations are the action of the Weyl group and we do not want to loose them. As shown in \cite{BS} the fermionic dualities also hold when the one loop fermionic nodes are deformed to those present in the all loop BS equations. The fermionic kernels we wrote are exactly the same as there and therefore the fermionic dualities are also present in our equations. On the other hand, the bosonic dualities studied in the $AdS_5/CFT_4$ in \cite{GV3} are much more restrictive on the form of the bosonic nodes. It seems hard to deform them away from their $2$-loop form while keeping the dualities valid. Notice indeed that the equation for the bosonic roots $u_2$ in (\ref{BAE}) is the same as for the $2$-loop Minahan-Zarembo equations (\ref{betheeqscfalt}). We will comment more on the importance of these dualities at the end of the next section and in appendix A.

\section{Strong coupling limit} \la{sec2}
In this section we analyze the strong coupling limit of the conjectured Bethe equations (\ref{BAE}).
We will explain how to encode them into a single ten-sheeted Riemann surface.
Then we will obtain a precise match between this surface and the algebraic curve recently proposed in \cite{curve}. This is an important check of our conjecture. In particular it shows
that all the quasi-classical results from the string theory side are
automatically incorporated into our equations.

To obtain the string classical limit the Bethe roots should scale as
$$
\sqrt{\lambda/2}\sim u_{a,j} \sim K_a \sim L \gg 1 \,,
$$
so that
$$
x^{\pm}=x\pm \frac{i}{2}\,\alpha(x)+\OO\(\frac1\lambda\)
$$
where
$$
\alpha(x)\equiv \frac{1}{\ff}\frac{x^2}{x^2-1} \,.
$$
This very same function was also introduced in \cite{curve}. To present the result of the expansion of the Bethe equations in the scaling limit, it is convenient to introduce the  resolvents
\beq
\nn G_a(x)=\sum_j\frac{\alpha(x_{a,j})}{x-x_{a,j}}\;\;,\;\;
 H_a(x)=\sum_j\frac{\alpha(x)}{x-x_{a,j}}\;\;,\;\;\bar
H_a(x)=H_a(1/x)\;\;,\;\;\bar G_a(x)=G_a(1/x)\,.
\eeq
In the limit $K_a\to \infty$ the roots $x_{a,j}$ condense into some cuts in the complex plane and these sums could be replaced by integrals using the densities of the Bethe roots \cite{C1,C2,C3,C4,C5,C6}.

First we expand the charges ${\cal Q}_n$ (\ref{charges}) in the scaling limit to find
\beq
G_{4}(x)+G_{\bar 4}(x) =- \sum_{n=0}^{\infty} \mathcal{Q}_{n+1} x^n\;,
\eeq
so that in particular, from (\ref{eq:E}) and (\ref{mom}), we have
\beqa
2\pi m&=&-G_4(0)-G_{\bar 4}(0)\;\;,\;\;E=-2 g \(G'_4(0)+G'_{\bar 4}(0)\)\;.
\eeqa
Now we have all the necessary ingredients to present the expansion of the Bethe ansatz equations (\ref{BAE}) in the classical scaling limit. We find
\beqa
2\pi n_{1}&=&-\frac{{\cal Q}_1+x{\cal Q}_2}{x^2-1}-H_2-\bar H_2+\bar H_4+\bar H_{\bar 4}\;\;,\;\;x\in {\cal C}_{1}\\
\la{exeq2}2\pi n_{2}&=&\quad\quad\quad\quad\,\,\;\quad-H_1+2\slashed H_2-H_3-\bar H_1+2\bar H_2-\bar H_3\;\;,\;\;x\in {\cal C}_{2}\\
2\pi n_{3}&=&+\frac{{\cal Q}_1+x{\cal Q}_2}{x^2-1}-H_2+H_4+H_{\bar 4}-\bar H_2\;\;,\;\;x\in {\cal C}_3\\
2\pi n_4&=&\frac{Lx/2g-{\cal Q}_1}{x^2-1}+H_3-2\slashed H_4+\bar H_1-\bar H_4+\bar H_{\bar 4}\;\;,\;\;x\in {\cal C}_4\; \\
2\pi n_{\bar 4}&=&\frac{Lx/2g-{\cal Q}_1}{x^2-1}+H_3-2\slashed H_{\bar 4}+\bar H_1-\bar H_4+\bar H_{\bar 4}\;\;,\;\;x\in {\cal C}_{\bar 4}\;.
\eeqa
where a slash stands for the average of the function above and below the cut.
To obtain these equations we first take the $\log$ of both sides of (\ref{BAE}) and obtain in this way the several $2\pi n_a$ coming from the various possible choices for the log branches. Furthermore, the products in (\ref{BAE}) become sums and in the scaling limit the summands can be expanded and simplified and we recognize the several resolvents introduced above.

Notice that to obtain these equations in the scaling limit the asymptotics (\ref{AFSl}) are important and so is the fact that the $u_4$ and $u_{\bar 4}$ roots in (\ref{BAE}) are coupled in the way indicated there. The fact that we will obtain a perfect match with the algebraic curve in \cite{curve} is therefore an important argument in favor of such structure in (\ref{BAE}). See also the next subsection and the discussion in section \ref{sec3} for further evidence.

Now the most important point: these equations for the resolvents  imply that the following five quasi-momenta
\beqa
&& q_3(x)=\frac{Lx/2g-{\cal Q}_1}{(x^2-1)}\,\,\,-H_1+\bar H_4+\bar H_{\bar 4}-\bar H_3\nn \,, \\
&& q_2(x)=\frac{Lx/2g+{\cal Q}_2x}{(x^2-1)}+H_2-H_1-\bar H_3+\bar H_2 \,, \nn\\
&& q_1(x)=\frac{Lx/2g+{\cal Q}_2x}{(x^2-1)}+H_3-H_2-\bar H_2+\bar H_1  \,, \la{qsH}\\
&& q_4(x)=\frac{Lx/2g-{\cal Q}_1}{(x^2-1)}\,\,\,-H_4-H_{\bar 4}+H_3+\bar H_1 \,, \nn\\
&& q_5(x)=\quad\quad\quad\quad\quad\quad \,\,+H_4-H_{\bar 4}+\bar H_4-\bar H_{\bar 4} \,, \nn
\eeqa
together with $\{-q_{1}(x),\dots,-q_{5}(x)\}$, form an algebraic curve. More precisely
$\{e^{ q_i},e^{-q_i}\}$ should be regarded as ten branches of the same analytical function.

To see this let us pick a pair of quasimomenta,  say $q_1$ and $q_2$, and consider the values of these functions immediately above and below a cut resulting from the condensation of a large number of $u_2$ Bethe roots. More precisely let us compute their discontinuity and average  for values of $x$ belonging to the cut.

By definition the discontinuities of $q_2$ and $-q_1$ on the cut ${\cal C}_{2}$ are equal (and proportional to the density of the $x_{2,j}$ roots)
\beq\la{delta}
q_2(x+i0)-q_2(x-i0)=-q_1(x-i0)+q_1(x+i0)\;\;,\;\;x\in {\cal C}_{2}\;.
\eeq
Then we notice that the equation \eq{exeq2} could be cast into
\beq\la{slash}
q_2(x+i0)+q_2(x-i0)=q_1(x+i0)+q_1(x-i0)+4\pi n_2\;\;,\;\;x\in {\cal C}_{2}\;.
\eeq
These two equations are equivalent to
\beq
e^{iq_2(x+i0)}=e^{iq_1(x-i0)}\;\;,\;\;e^{iq_2(x-i0)}=e^{iq_1(x+i0)}\;\;,\;\;x\in {\cal C}_{2}\;,
\eeq
which means that the functions $e^{iq_2(x)}$ and $e^{iq_1(x)}$ are glued to each other by the cut
${\cal C}_2$. Proceding like this for the several quasi-momenta we can see that the $10$ sheets are nicely glued together into a single Riemann surface.

This was the most nontrivial part. Now we simply need to identify the various analytical properties of the quasi-momenta defined above and check that they match precisely those appearing in the classical finite gap construction of \cite{curve}.

First notice that $q_1,q_2$ and $q_3,q_4$ are simply related by the inversion symmetry $x\to 1/x$  while $q_5$ is symmetric with respect to this transformation. Exactly the same behavior was found in the string algebraic curve \cite{curve}. The large $x$ asymptotics also match. For example, from the definition of $q_1$ we see that
\beq
q_1(x)\simeq \frac{L+E+K_3-K_2}{2g x}
\eeq
which is precisely the expression in \cite{curve}. Finally it is easy to see that the pole structure at $x=\pm 1$ is also as in \cite{curve}. In particular we observe the nontrivial synchronization of the residues of the several quasi-momenta which in \cite{curve} appear as a consequence of the string Virasoro constrains. For example, we expect the difference
\beq
q_1-q_4=\frac{{\cal Q}_2 x+{\cal Q}_1}{x^2-1}+H_4+H_{\bar 4}-H_2-\bar H_2
\eeq
to be regular at $x=\pm 1$
The combination $H_2+\bar H_2$ is regular at $x=\pm 1$ and the sum $H_4+H_{\bar 4}$ is directly related to the conserved charges ${\cal Q}_1$ and ${\cal Q}_2$ in a way that together with the first term we obtain a completely regular expression. The other quasi-momenta can be analyzed in a similar fashion.

This completes the comparison of all the analytical properties of the finite gap and Bethe ansatz quasi-momenta with a perfect match for every one of them.

\subsection{Semi-classical dressing factor} \la{HL}
To further check our conjecture let us compute the first leading correction to the AFS dressing kernel. We did this computation once in the context of the $AdS_5/CFT_4$ -- where the first quantum correction to the dressing kernel is given by the Henernadez-Lopez phase \cite{HL} -- in \cite{GV2} therefore we will be extremely brief now and omit the details. We refer to \cite{GV2} for further details.

The idea is that to get the 1-loop correction to the energy and the other local charges
 one should add zero point oscillations around the classical solution. Each mode of the oscillation corresponds to a particular small deformation of the classical solution. In the language of the algebraic curve each mode corresponds to a new pole with a tiny residue. More precisely
exciting a solution by a mode implies to add a fluctuation to a quasimomentum \cite{GV,vicedo}
\beq
\delta q(x)\sim \pm\frac{\alpha(x)}{x-x_n}\;.\la{extpoles}
\eeq
The fluctuations  have different polarizations which are summarized in the figure 2.
They correspond to different quasimomenta which should be excited. For example the first
excitation on the fig. 2. is labeled by $45$ which means that only $q_4$ and $q_5$
should have extra poles (\ref{extpoles}) when the classical solution is excited by that fluctuation. One should think about these poles as being a very small cut
connecting $q_4$ and $q_5$ therefore the residues should have opposite signs.
The points on the curve where one can open a small cut are some special loci where the sheets
of the curve are crossing each other. These points can be found from the equation
\beq
q_i(x_n)-q_j(x_n)=2\pi n\;. \la{map}
\eeq
The integer numbers $n$'s are called mode numbers and are the analogues of the Fourier mode numbers in the flat space.

\begin{figure}[t]
\centering \resizebox{80mm}{!}{\includegraphics{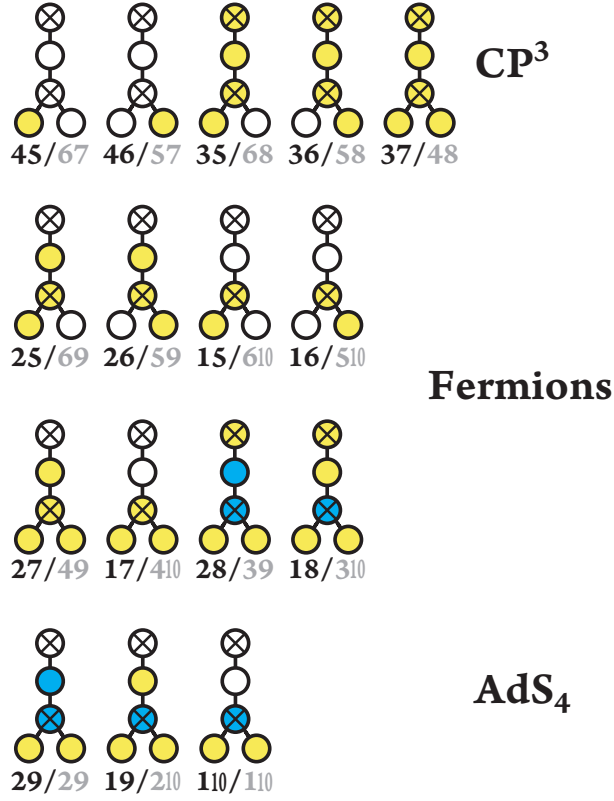}}
\caption{\small  The several states in the Hilbert space can be constructed in the usual oscillator representation. There is one oscillator per Dynkin node of the $OSp(2,2|6)$ super Dynkin diagram. A light (dark) gray shaded node corresponds to an oscillator excited once (twice). The number of times each oscillator is excited is the same as the number of Bethe roots of the corresponding type. } \label{excitations}
\end{figure}

Zero point oscillations correspond to the sum over halves of all possible fluctuations \cite{vicedo,GV}. In this way, the charges computed for these excited quasimomenta automatically gain a ground zero shift as predicted from field theory. We should therefore add each fluctuation in figure \ref{excitations} with a $1/2$ factor for bosonic excitations and a factor of $-1/2$ for the fermionic fluctuations and sum over the mode number $n$ of these fluctuations. For our goal we can replace sum over $n$ by an integral. Then we replace the variable $n$ by the position $x$ of the corresponding fluctuation using (\ref{map}) so that $dn=(q'_i-q'_j)dx/2\pi$.

Let us then look at the shift in the quasi-momenta $q_5=-q_6$.
From what we explained we see that there are only
4 bosonic fluctuations  $45,46,35,36$ and 4 fermionic ones $25,26,15,16$.
\beqa
\delta q_5&=&\frac{1}{2}\int \frac{dy}{2\pi} \frac{\alpha(x)}{x-y} \( (q_4'-q_5')-(q_4'-q_6')+(q_3'-q_5')-(q_3'-q_6')\)\\
&-&\frac{1}{2}\int \frac{dy}{2\pi} \frac{\alpha(x)}{x-y} \((q_2'-q_5')-(q_2'-q_6')+(q_1'-q_5') -(q_1'-q_6')\)\nn\\
\nn&+& (x\to 1/x)=0
\eeqa
Thus $q_5$ is not shifted.
Next let us consider the shift in $q_4=-q_7$. Again we should only consider fluctuations
from the fig. 2 with labels $4$ or $7$. We find three bosonic excitations $45,46,37$ and two fermionic poles $27,17$ so that
\beqa
\delta q_4&=&\frac{1}{2}\int \frac{dy}{2\pi} \frac{\alpha(x)}{x-y} \( -(q_4'-q_5')- (q_4'-q_6')-(q_3'-q_7')+(q_2'-q_7')+(q_1'-q_7')  \) \\
&-&\frac{1}{2}\int \frac{dy}{2\pi} \frac{\alpha(1/x)}{1/x-y} \( -(q_3'-q_5')-(q_3'-q_6')-(q_3'-q_7')+(q_2'-q_8')+(q_1'-q_8')\nn
\)\eeqa
The second line is added to satisfy the $x\to 1/x$ symmetry of the algebraic curve \cite{curve}, which relates $q_4$ and $q_3$. This expression can be simplified to
\beqa
\delta q_4&=&\frac{1}{2}\int \frac{dy}{2\pi} \(\frac{\alpha(x)}{x-y}-\frac{\alpha(1/x)}{1/x-y}\) \d_y\(q_1+q_2-q_3-q_4\)
\eeqa
using that from \eq{qsH}
\beq\nn
q_1+q_2-q_3-q_4=2\frac{{\cal Q}_1+x {\cal Q}_2}{x^2-1}+H_4+H_{\bar 4}-\bar H_4-\bar H_{\bar 4}={\cal Q}_1+G_4+G_{\bar 4}-\bar G_4-\bar G_{\bar 4} \,.
\eeq
Then, since $G_4+G_{\bar 4}=-\sum_{n=0}{\cal Q}_{n+1}y^n$, we get
\beqa\nn
\delta q_4(x)&=&\frac{1}{2}{\cal V}(x)\;\;,\;\;{\cal V}(x)=\alpha(x)\sum_{r=2,s>r}^{\infty}\frac{1+(-1)^{r+s}}{\pi}\frac{(r-1)(s-1)}{(s-r)(r+s-2)}\(\frac{{\cal Q}_r}{x^s}-\frac{{\cal Q}_s}{x^r}\)\;,
\eeqa
which in is precisely the Hernandez-Lopez phase \eq{crs}! Repeating the computation for $q_1$ and $q_2$ will lead to the same result.

Notice that all the quasimomenta $q_1,\dots,q_4$ are shifted by the same amount while $q_5$ is not shifted at all. This is precisely what ensures that these quantum fluctuations can be traded for the insertion of a dressing phase $e^{-i{\cal V}/2}$ in the momentum carrying nodes. Indeed, recall that each Bethe equation (\ref{BAE}) in the scaling limit is obtained by considering the difference of two consecutive quasi-momenta. The fact that all the first $4$ quasimomomenta are shifted by the same amount means the first three equations do not need to be modified. On the other hand the equations for the momentum carrying nodes which follow from $q^+_4-q_5^-=2\pi n$ and $q_4^++q_5^-=2\pi n$ are modified by the potential we just derived\footnote{In \cite{GV2} we obtained exactly the same expression $\frac{1}{2}\mathcal{V}$ in terms of the charges for the first four quasi-momenta there, $p_1,\dots,p_4$ while for the last four we obtained $-\frac{1}{2}\mathcal{V}$. Therefore when we considered the difference of consecutive quasimomenta we concluded that only the middle node was modified and the modification was $\frac{1}{2}\mathcal{V}-\(-\frac{1}{2}\mathcal{V}\)$ which is precisely twice as much as here, precisely in agreement with the discussion in the introduction.}. Recall moreover that according to (\ref{HLQ}) the correction we found points again towards an interaction between the roots $u_4$ and $u_{\bar 4}$ as announced in (\ref{BAE}). See the previous subsection as well as section \ref{sec3} for further evidence for this structure.

Notice that for all the argument to go through it is important that the function $\ff$ mentioned in the introduction contains no $\mathcal{O}(1)$ term in the large $\lambda$ expansion as mentioned in the introduction.

Let us come back to the dualities in the Bethe ansatz equations mentioned in section \ref{sec1}.
The $1/\sqrt{\lambda}$ effects present in the string semi-classical quantization are accounted by the Hernandez-Lopez dressing phase. But there is also another type of correction which appears at the same order and must also be taken into account to obtain a complete agreement between the BAE (\ref{BAE}) and the string semiclassical spectrum. These corrections are the ones appearing from the finite size corrections errors introduced when going to the continuum limit and expanding the roots in the scaling limit. In \cite{GV3} we showed that these corrections were precisely of the required form to ensure the proper match with the string semi-classical quantization around any classical configuration. To be able to derive this it was crucial that the dualities mentioned in the previous section existed and it seems to us very unlikely that without such dualities a match would take place.

\section{$SU(2|2)$ symmetry and the dressing kernel} \la{sec3}
As shown in the first two lines of figure \ref{excitations}, there are four fluctuations where $u_{\bar 4} $ roots are not excited and four fluctuations where $u_4$ roots are not excited. These are the two $\textbf{4}$ dimensional short representations of the two $SU(2|2)$ subsectors found in \cite{MZ,BMN2}.
As explained in \cite{BMN2} each $SU(2|2)$ sector is centrally extended in the exact same way as observed in the $AdS_5/CFT_4$ duality by Beisert \cite{B22} (see also \cite{Arutyunov:2006ak}). The value of the central charge is related to $\ff$ mentioned in the introduction and this fixes the dispersion relation to be of the form (\ref{Ep}) \cite{B22,BMN2}.

Moreover, as explained in \cite{B22} the Bethe equations for a system with $SU(2|2)$ extended symmetry are completely fixed up to a scalar factor. This is indeed built in our conjectured equations (\ref{BAE}). Namely, if we focus on one of the $SU(2|2)$ sub-sectors -- by considering for example no $u_{\bar 4}$ roots --  equations (\ref{BAE}) reduce to the Bethe equations for a $SU(2|2)$ extended symmetric system \cite{B22}. Equations (\ref{BAE}) are the most natural way to combine the $SU(2|2)\oplus SU(2|2)$ symmetry in a $OSp(2,2|6)$ symmetric system of Bethe equations. Figure \ref{excitations} also seems to indicate that the remaining $8$ fluctuations organize into a $(4|4)$ multiplet as argued in \cite{BMN2}.

Note that since we deal with a $SU(2|2)\oplus SU(2|2)$ symmetry, rather than the $SU(2|2)^2$ in the Beisert-Staudacher equations \cite{BS,B22}, we have two momentum carrying nodes (for $u_4$ and for $u_{\bar 4}$) each of them with its own dressing Kernel. The dressing kernel for each node is moreover the square root of the dressing kernel present in the $AdS_5/CFT_4$ duality. This is probably also a consequence of the symmetries of the theory. As recalled by Janik and Lukowski in \cite{Janik:2008hs} there is usually a strong relation between the dressed part of the Bethe equations and the dressing kernel of the momentum carrying nodes. The idea is that in integrable field theories the scalar factor can be expressed as a convolution of simple kernels appearing in the nested levels of the Bethe ansatz equations. This is of course an empirical observation but so far it seems to work.  In our nested Bethe equations we have two momentum carrying nodes. On top of each of these there is a $SU(2|2)$ tower -- see figure \ref{BAEpic}. On the other hand, in the Beisert-Staudacher equations, we have a single momentum carrying node connected with two $SU(2|2)$ wings. Thus it is very natural to expect that the dressing phase for each of our momentum carrying roots is simply half of that obtained in the $AdS_5/CFT_4$ duality in \cite{BES}.

In (\ref{BAE}) the momentum carrying roots $u_4$ and $u_{\bar 4}$ are also connected by a BES dressing kernel \cite{BES} which takes exactly the same form as the kernel appearing in the self-interaction of the momentum carrying roots, see figure \ref{BAEpic}. We already found evidence for the precise structure in the previous sections where we studied the strong coupling limit of the Bethe equations. Here we will argue from another point of view why this structure is to be expected.

First we need to recall the observation of \cite{GK}. In this paper, following \cite{GKSV},  integrable relativistic $SO(n)$ sigma models were considered. The quantization of such models is obtained by solving a  set of Bethe equations which quantize the physical momenta of the relativistic particles and the isotopic momenta of the $SO(n)$ spin waves. In \cite{GK} it was understood how to eliminate the physical momenta from these Bethe equations to obtain an effective equation for the spin isotopic degrees of freedom. In the classical limit, the effective equations were then seen to match not only the classical algebraic curves in \cite{C1,C3} but also the conjectured string Bethe equations \cite{AFS} studied in the context of integrability in $AdS_5/CFT_4$. This was possible because in the classical limit the string motion can be consistently truncated to a $S^n$ subspace in $AdS_5\times S^5$.  It is clear that we can learn a lot about the structure of the strong coupling limit of the  $AdS/CFT$ Bethe equations from these simpler relativistic toy models. For example, to learn about the structure of the $SU(2)\times SU(2)$ closed sector of the ABJM Chern-Simons theory theory, recently studied in \cite{BMN3}, a nice toy model is the $SU(2)$ principal chiral field  whose symmetry is $SU(2)_L\times SU(2)_R$. The spectrum of this model is found from solving the Nested Bethe equations \cite{Z1,Z2}
 \begin{eqnarray}
e^{-i \mathcal{L} p(\theta_\alpha) }&=& \prod_{\beta\neq \alpha}\, S_0^{\,2} \(
\theta_\alpha-\theta_\beta \) \prod_j\frac{\theta_\alpha-u_{4,j}+i/2}{\theta_\alpha-u_{4,j}-i/2}\,
\prod_k\frac{\theta_\alpha-u_{\bar 4,k}+i/2}{\theta_\alpha-u_{\bar 4,k}-i/2}\,, \label{DBAE1} \\
1&=&\prod_\beta\frac{u_{4,j}-\theta_\beta-i/2}{u_{4,j}-\theta_\beta+i/2}
\prod_{i\neq j} \frac{u_{ 4,j}-u_{4,i}+i}{u_{ 4,j}-u_{4,i}-i}\,,  \label{DBAE2}\\
1&=&\prod_\beta\frac{u_{\bar 4,k}-\theta_\beta-i/2}{u_{\bar 4,k}-\theta_\beta+i/2} \prod_{l\neq k}
\frac{u_{\bar 4,k}-u_{\bar 4,l}+i}{u_{\bar 4,k}-u_{\bar 4,l}-i}\,, \label{DBAE3}
\end{eqnarray}
where $S_0$ is a known function whose explicit form is not relevant for our discussion. We can now understand what happens when we find the position of the Bethe roots $\theta_\alpha$ from the first equation and plug them into the last two equations (\ref{DBAE2},\ref{DBAE3}). We will obtain in this way two effective equations for the two spin rapidities associated with each of the two $SU(2)$ symmetry groups. Without any computation at all it is already possible to learn a lot about the symmetry of the obtained equations. In the classical limit we take the log of these equations and transform the obtained sums into integrals. The density of $\theta$ roots obtained by solving the first (integral) equation is clearly of the form
\beq
\rho(\theta)=\rho_0(\theta)+\sum_{j=1}^{K_4} \rho_1(\theta,u_{4,j})+\sum_{j=1}^{K_{\bar 4}} \rho_1(\theta,u_{\bar4, j}) \la{rho} \,.
\eeq
Notice that there is a single function $\rho_1$ in the last two terms. This follows trivially from the fact that the $u_4$ and $u_{\bar 4}$ enter in (\ref{DBAE1}) in equal footing but will have important consequences.  When we integrate out the physical rapidities in each of the products in the last two Bethe equations (\ref{DBAE2},\ref{DBAE3}) using $\rho(\theta)$, we will generically find
 \begin{eqnarray}
1&=&F(u_{4,j}) \prod_{i=1}^{K_4} \sigma(u_{4,j},u_{4,i}) \prod_{i=1}^{K_{\bar 4}} \sigma(u_{4,j},u_{{\bar 4},i}) \prod_{i\neq j}^{K_4} \frac{u_{ 4,j}-u_{4,i}+i}{u_{ 4,j}-u_{4,i}-i}\,,  \nn \\
1&=&F(u_{\bar 4,j})\prod_{i=1}^{K_4} \sigma(u_{\bar 4,j},u_{ 4,i}) \prod_{i=1}^{K_{\bar 4}} \sigma(u_{4,j},u_{{\bar 4},i})
 \prod_{i\neq j}^{K_{\bar 4}} \frac{u_{\bar 4,k}-u_{\bar 4,l}+i}{u_{\bar 4,k}-u_{\bar 4,l}-i}\,. \nn
\end{eqnarray}
The first term $F(u)$ comes from the contribution from $\rho_0$ and the first two products in each line come from the convolutions with the last two terms with $\rho_1$ in (\ref{rho}). $F(u)$ should be thought of as $e^{i L p_{eff}(u)}$, an effective dispersion relation for the $SU(2)$ magnons. As for $\sigma$, it appears as a new effective self-interaction that emerges in the  $u_4$ and $u_{\bar 4}$ equations but also as a kernel coupling the two $SU(2)$! The coupling kernel and the new self interactions are therefore, by simple symmetry arguments, exactly the same! This shows that the structure we propose for the interaction of the momentum carrying nodes of the $AdS_3/CFT4$ Bethe equations is not exotic at all but rather what we would expect.

This is even more so when we recall \cite{GK} that, when this procedure is carried in detail in the classical limit, the effective dispersion relations which appear are exactly the $\(x^+/x^-\)^L$ appearing in our equations and, moreover, the kernels $\sigma$ are precisely of the AFS form! The AFS kernel \cite{AFS} seems to be  highly universal and it is nice to see that it fits neatly in our conjectured equations.

\section*{Acknowledgments}
We would like to thank N.~Dorey, V.~Kazakov, I.~Kostov,  J~Penedones, D.~Serban, I.~Shenderovich, D.~Volin, K.~Zarembo and specially J.~Minahan for many useful discussions and insightful comments. We would like to thank T.~Vieira for a thorough reading of the manuscript. PV is funded by the Funda\c{c}\~ao para a Ci\^encia e Tecnologia fellowship {SFRH/BD/17959/2004/0WA9}. NG was partially supported by RSGSS-1124.2003.2, by RFFI project grant 06-02-16786 and ANR grant INT-AdS/CFT (contract ANR36ADSCSTZ).
NG would like to thank les Houches where parts of the work were done for hospitality.
%%%%%%%%%%%%%%%%%%%%%%%%%%%%%%%%%%%%%
%%%%%%%%%%%%%%%%%%%%%%%%%%%%%%%%%%%%%
%%%%%%%%%%%%%%%%%%%%%%%%%%%%%%%%%%%%%

\section*{Appendix A -- Fermionic dualities}
\begin{figure}[t]
\centering \resizebox{140mm}{!}{\includegraphics{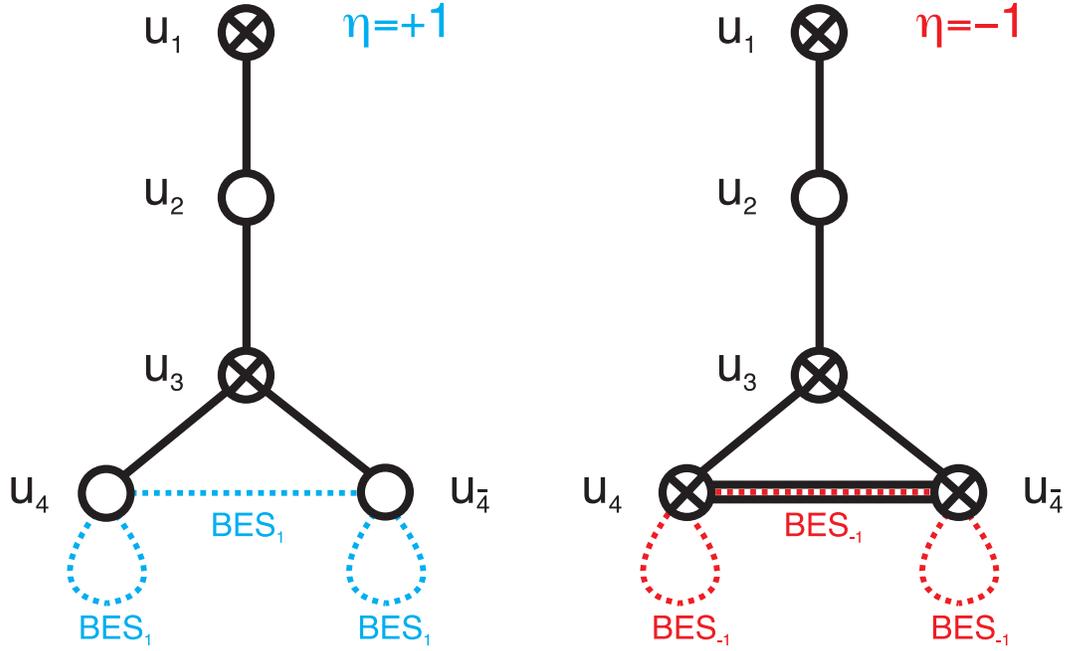}}
\caption{\small  Bethe equations for the two choices of the grading $\eta=\pm 1$. The subscripts $\pm 1$ in the dashed lines correspond to the kernels $\sigma_{BES}(u,v)$ for $\eta=1$ and to a kernel $\sigma_{BES}(u,v) \frac{x^-(u) -x^+(v)}{x^+(u)-x^-(v)}$ for $\eta=-1$.
} \label{BAEpiceta}
\end{figure}

We can transform our equations (\ref{BAE}) into an equivalent set of Bethe equations by application of the fermionic dualities. This follows \cite{BS} closely. We construct the polynomial
\beqa
P(x)&=&\prod_{j=1}^{K_4}(x-x_{4,j}^+)\prod_{j=1}^{K_{\bar 4}}(x-x_{\bar 4,j}^+)\prod_{j=1}^{K_2}(x-x_{2,j}^-)(x-1/x_{2,j}^-) \nn \\
&-&\prod_{j=1}^{K_4}(x-x_{4,j}^-)\prod_{j=1}^{K_{\bar 4}}(x-x_{\bar 4,j}^-)\prod_{j=1}^{K_2}(x-x_{2,j}^+)(x-1/x_{2,j}^+)  \la{P1}
\eeqa
and by virtue of the Bethe equations (\ref{BAE}) for the fermionic roots $u_1$ and $u_3$ we find that this polynomial has zeros for $x=x_{3,j}$ and for $x=1/x_{1,j}$ so that
\beqa
P(x)&=&\prod_{j=1}^{K_3}(x-x_{3,j})\prod_{j=1}^{K_1}(x-1/x_{1,j})\prod_{j=1}^{\t K_3}(x-\t x_{3,j})\prod_{j=1}^{\t K_1}(x-1/\t x_{1,j}) \la{P2}
\eeqa
where the $\t x$ are the remaining zeros of the polynomial. Then by equating (\ref{P1}) and (\ref{P2}) and evaluating this relation at some particular values like $x_{4,k}^+$, $x_{\bar 4,k}^-$, $x_{3,k}^-$ etc (see \cite{BS} for details) we find several relations between the tilded $\t x_1$ and $\t x_3$ and the original Bethe roots $x_1$ and $x_3$. In particular we see that the old Bethe roots $x_1$ and $x_3$ can be replaced by their tilded counterparts provided we modiify the Bethe equations to
\begin{eqnarray}
1&=&
\prod_{j=1}^{K_2}
\frac{u_{1,k}-u_{2,j}+\frac{i}{2} }{u_{1,k}-u_{2,j}-\frac{i}{2} }
\prod_{j=1}^{K_{4}}
\frac{1-1/x_{1,k} x^+_{4,j}}{1-1/x_{1,k}x_{4,j}^-}
\prod_{j=1}^{K_{\bar 4}}
\frac{1-1/x_{1,k} x^+_{\bar 4,j}}{1-1/x_{1,k}x_{\bar 4,j}^-} \,,
 \nn \\
\nn 1&=&
\prod_{j\neq k}^{K_2}
\frac{u_{2,k}-u_{2,j}-i }{u_{2,k}-u_{2,j}+i }
\prod_{j=1}^{K_1}
\frac{u_{2,k}-u_{1,j}+\frac{i}{2} }{u_{2,k}-u_{1,j}-\frac{i}{2} }
\prod_{j=1}^{K_3}
\frac{u_{1,k}-u_{3,j}+\frac{i}{2} }{u_{1,k}-u_{3,j}-\frac{i}{2} }\,,
\\
\nn 1&=&
\prod_{j=1}^{K_2}
\frac{u_{3,k}-u_{2,j}+\frac{i}{2} }{u_{3,k}-u_{2,j}-\frac{i}{2} }
\prod_{j=1}^{K_4}
\frac{x_{3,k} -x^+_{4,j}}{x_{3,k} -x^-_{4,j}}
\prod_{j=1}^{K_{\bar 4}}
\frac{x_{3,k} -x^+_{\bar 4,j}}{x_{3,k} -x^-_{\bar 4,j}} \,
\\
\la{BAEeta} \(\frac{x^+_{4,k}}{x^-_{4,k}}\)^{
L} &=&
\prod_{j\neq k}^{K_4}
\frac{u_{4,k}-u_{4,j}+i}{u_{4,k}-u_{4,j}-i}  \,
\prod_{j=1}^{K_1}
\(\frac{1-1/x^-_{ 4,k} x_{1,j}}{1-1/x^+_{ 4,k} x_{1,j}} \)^{\eta}
\prod_{j=1}^{K_3}
\( \frac{x^-_{ 4,k}-x_{3,j} }{x^+_{ 4,k}-x_{3,j}} \)^{\eta}   \times
 \\
&\times &\prod_{j\neq k}^{K_{ 4}}
\sigma_{BES}(u_{  4,k},u_{  4,j}) \(\frac{x_{4,k}^--x_{4,j}^+}{x_{4,k}^+-x_{4,j}^-} \)^{\frac{1-\eta}{2}} \prod_{j=1}^{K_{  \bar 4}} \sigma_{BES}(u_{ 4,k},u_{ \bar 4,j})  \(\frac{x_{4,k}^- -x_{\bar 4,j}^+}{x_{4,k}^+-x_{\bar 4,j}^-} \)^{\frac{1-\eta}{2}}\nn \,,\\
\(\frac{x^+_{\bar 4,k}}{x^-_{\bar  4,k}}\)^{
L} &=&
\prod_{j\neq k}^{K_{\bar 4}}
\frac{u_{\bar  4,k}-u_{\bar 4,j}+i}{u_{\bar 4,k}-u_{\bar 4,j}-i}  \,
\prod_{j=1}^{K_1}
\(\frac{1-1/x^-_{\bar 4,k} x_{1,j}}{1-1/x^+_{\bar 4,k} x_{1,j}} \)^{\eta}
\prod_{j=1}^{K_3}
\( \frac{x^-_{\bar 4,k}-x_{3,j} }{x^+_{\bar 4,k}-x_{3,j}} \)^{\eta}   \times
\nn \\
&\times &\prod_{j\neq k}^{K_{\bar 4}}
\sigma_{BES}(u_{ \bar 4,k},u_{ \bar 4,j}) \(\frac{x_{\bar 4,k}^--x_{\bar 4,j}^+}{x_{\bar 4,k}^+-x_{\bar 4,j}^-} \)^{\frac{1-\eta}{2}}
\prod_{j=1}^{K_{  4}} \sigma_{BES}(u_{ \bar 4,k},u_{ 4,j})  \(\frac{x_{4,k}^- -x_{\bar 4,j}^+}{x_{4,k}^+-x_{\bar 4,j}^-} \)^{\frac{1-\eta}{2}}\nn \,,
\end{eqnarray}
where $\eta=-1$. We removed the tilde's from all the $\t x_1$ and $\t x_3$. Equations (\ref{BAEeta}) for the gradings $\eta=\pm 1$ are two dual ways of writing the same Bethe equations. They correspond to different choice of Dynkin diagrams which, we recall, for superalgebraic is not unique. The two possible structures are presented in figure \ref{BAEpiceta}.

\subsection*{Appendix A.1 -- Scaling function}
In this section we will explain that there is a particular configuration of Bethe roots governed by absolutely the same equations as obtained in the $AdS_5/CFT_4$ duality. It is useful to work with the dualized equations (\ref{BAEeta}) with $\eta=-1$. Then we consider a configuration of Bethe roots with the same number of $u_4$ and $\bar u_4$ roots whose positions we take to be the same,
\beq
u_{4,k}=u_{\bar 4,k}\equiv u_k \,\,\, , \,\,\, k=1,\dots,S \gg 1 \,,
\eeq
and no auxiliary roots. Then both momentum carrying nodes yield identical equations for the positions $u_k$,
\begin{eqnarray}
\(\frac{x^+_{k}}{x^-_{ k}}\)^{
L} &=&
-\prod_{j\neq k}^{S}
\frac{u_{k}-u_{j}+i}{u_{k}-u_{j}-i}  \(\frac{x_{k}^- -x_{j}^+}{x_{k}^+-x_{j}^-} \)^{2}
\sigma_{BES}^2(u_{ k},u_{ j})\,. \la{sl2}
\end{eqnarray}
Notice that the kernel becomes squared so that these equations are precisely the $SL(2)$ BAE appearing in the $AdS_5/CFT_4$ Beisert-Staudacher equations except for the extra minus sign in the r.h.s!
In particular they are the starting point for the construction of the Eden-Staudacher \cite{ES}, Beisert-Eden-Staudacher \cite{BES} and Freyhult-Rej-Staudacher equations \cite{FRS} where the scaling function is studied thoroughly. The minus sign will mean that the mode numbers corresponding to the ground state will not be $n=\pm 1$ but rather $n=\pm 1/2$ and this will halve all results.
Therefore we conclude that  -- modulo the replacement of $\sqrt{\lambda}$ by $4\pi \ff$ and the division by $2$ as explained in the introduction -- the scaling function is precisely the same! For the generalized scaling function we obtain\footnote{We thank D.Volin for a discussion  on this point.}
\beq
f_{CS}\(\lambda,\frac{J}{\log(S)}\)=\frac{1}{2}\,f_{N=4}\(\lambda,\frac{2J}{\log(S)}\)_{\frac{\sqrt{\lambda}}{4\pi}\to h(\lambda) }\,.
\eeq

In the dual version of the Bethe equations we obtained the bosonic $SL(2)$ sector by considering pairs of excitations in the two fermionic nodes. In the original grading represented in fig. \ref{BAEpic} this amounts to taking $K_4=K_{\bar 4}=2K_3$ which from figure \ref{excitations} is clearly seen to have the appropriate quantum numbers to be called an SL(2) sector.

In the weak coupling limit from (\ref{sl2}) and (\ref{eq:E}) we find
\begin{eqnarray}
\(\frac{u+i/2}{u-i/2}\)^{L} &=&
-\prod_{j\neq k}^{S}
\frac{u_{k}-u_{j}-i}{u_{k}-u_{j}+i}  \,\, , \,\, E=\sum \frac{2\lambda^2}{u_j^2+\frac{1}{4}}
\end{eqnarray}
so that, from \cite{ES}, we find
\beq
E=4\lambda^2 \log S \,.
\eeq
where the $4$ prefactor instead of the $8$ comes from the half-integer mode numbers as explained above. Notice that to obtain this weak coupling result we need not use our conjectured equations at all. We could simply use the Minahan-Zarembo 2-loop equations to arrive at precisely the same result. Comparing with \cite{Aharony:2008ug} (equation 4.17) we see that there is a mismatch by a factor of $4$ which would be interesting to understand.

\end{document}